# Moiré-induced Vibrational Coupling in Double Walled Carbon Nanotubes

Georgy Gordeev[1*], Sören Waßerroth[1,2], Han Li[3], Benjamin Flavel[3], & Stephanie Reich[1]

1. Department of Physics, Freie Universität Berlin, Arnimallee 14, 14195 Berlin, Germany
2. Fritz-Haber-Institut der Max-Planck-Gesellschaft, Faradayweg 4-6, 14195 Berlin
3. Institute of Nanotechnology, Karlsruhe Institute of Technology, Karlsruhe, Germany

[*] gordeev@zedat.fu-berlin.de

**Abstract**

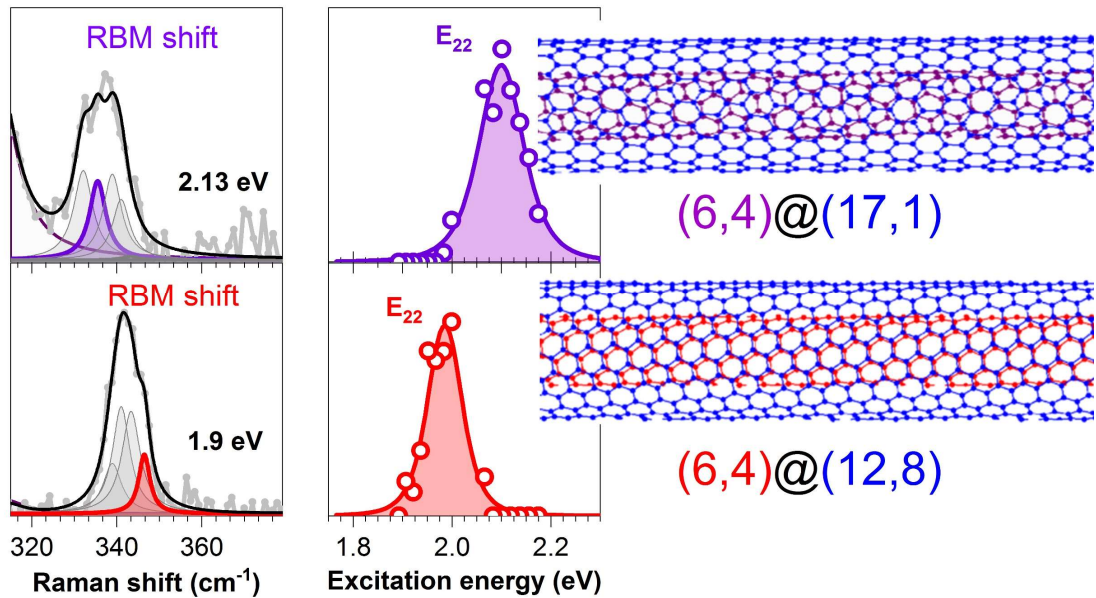

Moiré patterns are additional, long-range periodicities in twisted crystalline bilayers. They are known to fundamentally change the electronic states of the layers, but similar effects on their mechanical and vibrational properties have not been discussed so far. Here we show that the Moiré potential shifts the radial breathing mode in double walled carbon nanotubes (DWCNTs). It is expected that the frequency change is proportional to the shift in optical transition energies, which are induced by the Moiré patterns. To verify our model we performed resonant Raman scattering on purified and sorted semiconducting DWCNTs. We find that the radial breathing mode shifts up to 14 cm$^{-1}$ higher in energy followed by optical transitions energies displacement up to 200 meV to lower energies, compared to the single-walled tubes. We show how to identify the strong coupling condition in DWCNTs from their phonon frequencies and construct a Kataura plot to aid their future experimental assignment.

Introduction

Moiré patterns arise from the overlap of two displaced and identical crystalline lattices[1], they have been shown to drastically affect the electronic properties of two-dimensional materials. For example, graphene bilayers with small twist angles show correlated electronic behaviour with insulating and superconducting phases[2] and additional van Hove singularities emerge for larger angles[3]. In transition metal dichalcogenides moiré patterns may reconstruct the lattice, resulting in piezoelectric domains[4]. If two chemically different layers are combined, interlayer excitons form where one layer contains an electron and the other layer a hole. The lifetime of these quasiparticles depends on the twist angle and moiré structure[5].

Moiré patterns in one-dimensional (1D) systems[6] received less attention than their two-dimensional counterparts, although, once formed, they will be extremely stable and the 1D confinement introduces an additional degree of freedom for manipulating states[7–9]. Double-walled carbon nanotubes (DWCNT)[10] have distinct moiré patterns that arise from the two concentrically aligned single walled carbon nanotubes with different twist along their axis[7]. Depending on the chirality of the two tubes, the band gap shrinks due to the moiré interaction [8,9] or insulating flat bands are expected[7], very similar to the ones found in bilayer graphene[2]. The first challenge in the study of DWCNTs is that there are many possible structures and combinations that are associated with different electronic properties. Nanotube properties depend on diameter, chiral angle, and handedness, which are all determined by their $(n_1, n_2)$ chirality[11,12]. Single-walled nanotubes (SWCNTs) are either metallic (M) or semiconducting (S) depending on their chirality[11] and this results in four basic configurations of inner@outer nanotubes for DWCNTs: S@S, M@M, M@S, S@M[13]. Since as grown raw soot usually contains all four types and the wall-to-wall coupling depends sensitively on their electronic character, structure-specific studies on bulk DWCNT samples have been very challenging so far [14,15]. In a recent work we succeeded in separating DWCNTs by their electronic type[16–18]. This now enables the study of their properties and moiré physics in a wide range of DWCNTs.

Despite the key role that moiré effects play for the electronic states, the vibrational degrees of freedom have been examined rarely within moiré physics. Moiré patterns activate phonon modes that fulfil the selection rule of a vanishing wave vector only for the superstructure[19]. The moiré period also affects interlayer spacing and, thereby, indirectly changes the frequencies of the interlayer breathing modes by a few wavenumbers[20]. The vibrations of DWCNTs, however, were analysed from a purely mechanical point of view[15,21–25]. The phonon frequencies of DWCNTs split and shift compared to their SWCNT counterparts, which was explained as an effective strain that varies with wall-to-wall distance[21,23,26–28].

In this letter, we show that moiré coupling determines the vibrational properties of DWCNTs. First, a moiré-based model is proposed for phonons with radial eigenvectors. The phonon frequencies increase with moiré coupling strength. We confirm the predictions by resonance Raman spectroscopy of S@S enriched DWCNTs. The electronic transition energies get smaller,[7] while the frequency of RBM phonons increase in excellent agreement with our moiré-based theory. Finally, we construct a Kataura plot for semiconducting DWCNTs to simplify their future assignment.

**Results**

The atomic structure of two concentric CNTs and their resulting moiré patterns determine the coupling between the walls of a double walled tube. A DWCNT is characterized by four indices $(n_1^i, n_2^i)@(n_1^o, n_2^o)$ that specify the chiral indices of the inner (*i*) and outer (*o*) SWCNT. These indices produce different moiré patterns, see Fig. 1a. We will discuss the moiré physics by exemplarily comparing the (6,4)@(12,8) with the (6,4)@(17,1) DWCNTs. One expects the clear superstructure of the (6,4)@(12,8) DWCNT to show stronger interwall coupling than the (6,4)@(17,1) tube. Indeed, the coupling is stronger for two tubes with similar chiral angle, i.e., for small angles $\varphi\angle(\boldsymbol{C^i}, \boldsymbol{C^o})$ between the two chiral vectors, as is the case for the (6,4)@(12,8) tube[7]. $\boldsymbol{C^i} = n_1^i \boldsymbol{a_1} + n_2^i \boldsymbol{a_2}$ corresponds to the inner and $\boldsymbol{C^o} = n_1^o \boldsymbol{a_1} + n_2^o \boldsymbol{a_2}$ to the outer wall, with $\boldsymbol{a_1}$ and $\boldsymbol{a_1}$ being the basis vectors of graphene[12]. The chiral vectors determine the periodicity of the moiré superstructure.

The coupling strength of a DWCNT can be deduced from its structure in reciprocal space. The Brillouin zone (BZ) of DWCNT is obtained from the inner tube plus the BZ of the outer tube submitted to two symmetry operations[7], see Fig. 1b. First, we rotate the BZ by $\varphi$, this operation $P$ aligns $\boldsymbol{C^i}$ and $\boldsymbol{C^o}$. For inner and outer tubes with opposite handedness, twice chiral angle of the outer wall is added to $\varphi$. Second, we apply a uniaxial expansion $S^{-1}$. Its inverse operation $S$ describes a uniaxial contraction of the Bravais lattice along $\boldsymbol{C^i}$ by $\frac{|\boldsymbol{C^i}|}{|\boldsymbol{C^o}|}$, which scales the two chiral vectors to the same length. The reciprocal basis vectors of the resulting moiré superstructure are $\boldsymbol{G_l^M} = (\mathbb{1} - S^{-1}P^{-1})\boldsymbol{b_l}$ ($l = 1,2$), where $\mathbb{1}$ is the identity matrix and $\boldsymbol{b_l}$ are the reciprocal basis vectors of graphene. In the (6,4)@(12,8) tube $\boldsymbol{G_1^M}$ almost connects the two K point of the BZ of the inner and outer wall, Fig. 1b. Such a DWCNT is considered to be in the strong regime of chiral moiré coupling; the same applies if $\boldsymbol{G_2^M}$ or their sum $\sum \boldsymbol{G_l^M}$ connects the two K points. This condition can be expressed geometrically. It requires the two angles $\vartheta\angle(\boldsymbol{C^i} - \boldsymbol{C^o}, \boldsymbol{a_1} + \boldsymbol{a_2})$ and $\varphi$ to be close to zero[7]. Figure 1a illustrates this for three different DWCNTs, where $\varphi = 0$ and $\vartheta = 6.6°$ in the (6,4)@(12,8) tube, ($\varphi = 3.3°$, $\vartheta = 0°$) in the (6,4)@(11,9) and ($\varphi = 20.5°$, $\vartheta = 45°$) in the (6,4)@(17,1) DWCNTs.

The moiré pattern changes the electronic band structure of DWCNTs compared to the individual nanotubes. The Hamiltonian $H_{i,o}$ of a DWCNT can be expressed using the model developed by *Koshino et al.* as [7]:

$$H_{i,o} = \begin{pmatrix} H_i(\boldsymbol{k}) & U^\dagger \\ U & H_o(\boldsymbol{k}) \end{pmatrix}, \qquad (1)$$

where U is interlayer coupling matrix element, $H_i(\boldsymbol{k})$ and $H_o(\boldsymbol{k})$ are Hamiltonians of inner and outer wall, respectively. We use the tight-binding model of graphene for $H_i(\boldsymbol{k})$, with hopping energy $y_c = 2.7$ eV[12,29]. $H_o(\boldsymbol{k})$ is obtained from the tight-binding Hamiltonian by transforming the vectors connecting neighboring atoms according to $T_u^0 = T_u^i SP$, see Supplementary. The interlayer matrix element is expressed in terms of the moiré vectors as[7]:

$$U = u_0(R)\left(\mathbb{1} + \begin{pmatrix} 1 & \omega^{-1} \\ \omega^1 & 1 \end{pmatrix} e^{I\xi \boldsymbol{G_1^M} \cdot \boldsymbol{r}} + \begin{pmatrix} 1 & \omega^1 \\ \omega^{-1} & 1 \end{pmatrix} e^{I\xi(\boldsymbol{G_1^M}+\boldsymbol{G_2^M})\cdot \boldsymbol{r}}\right), \qquad (2)$$

where $\omega = exp(2\pi I\xi/3)$, and $\xi = \pm 1$ denotes time-reversal partners. $u_0(R)$ is a coupling amplitude depending on the interlayer distance. The band structure can be calculated numerically by the zone folding approach. We use Eq. (1) with the quantization condition around the circumference of $\mathbf{k} \cdot \mathbf{C}^i = 2\pi m$, where $m$ enumerates the cutting line, Fig. 1b.

The electronic dispersion of the SWCNT gets perturbed by the formation of a DWCNT. The band structure of the (6,4)@(12,8) nanotube is depicted in Figure 1d. In the uncoupled case, the $E_{22}$ band energies from inner and outer walls overlap for this particular pair of chiralities. In the coupled case the position of the $E_{22}$ electronic bands from the (12,8) outer wall move apart, increasing the second band gap. For the (6,4) inner wall the energies move closer together reducing the band gap. Figure 1c schematically shows coupling conditions for all realistic semiconducting outer walls of a (6,4) inner tube with interwall spacing 0.26-0.41 nm. The electronic bands perturb stronger for small $\vartheta$, $\varphi$, and inter-wall distances.

To study the effect of moiré coupling on the vibrational frequencies we focus on the radial breathing mode (RBM) with an eigenvector in the radial direction[12]. For SWCNTs the RBM frequency scales with ~1/d, (where d is the tube diameter)[12] and this has been widely used to study nanotube interactions with various exterior[30–32] and interior[33,34] environments. The moiré coupling changes the total energy of the inner tube by

$$\Delta E(R)_i = \frac{1}{|\mathbf{k}_\parallel|} \sum_m \int (E_{DW}^i(\mathbf{k}) - E_{SW}^i(\mathbf{k})) d\mathbf{k} \qquad (3)$$

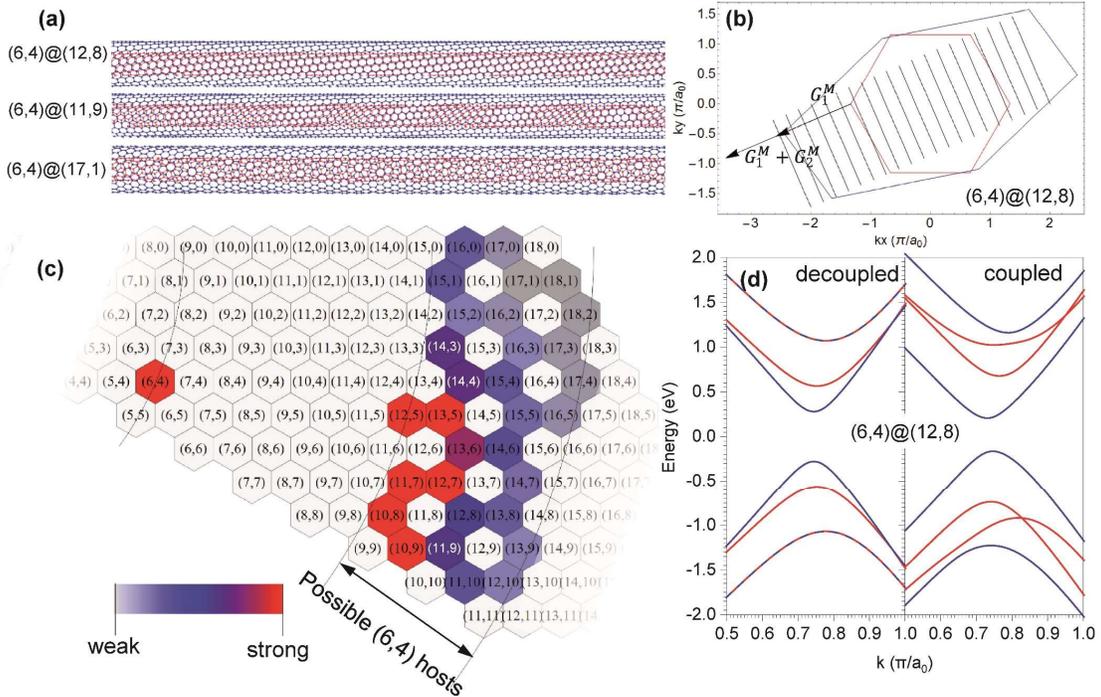

**Figure 1.** (a) Moiré induced effects in DWCNTs consisting of a (6,4) inner-wall (red) with different outer-walls (blue). (b) The Brillouin zone of the (6,4)@(12,8) DWCNT with moiré vectors included, $a_0$=0.246 nm. The cutting lines are translated into the first BZ (c) Possible outer-wall semiconducting nanotubes for a (6,4) inner-wall plotted on a model graphene sheet, where the color indicates the variation in coupling strength from weak (gray) to strong (red). (d) Band structure of a (6,4)@(12,8) nanotube with (right) and without (left) coupling. Bands originating from the (6,4) wall are shown in red, bands of the (12,8) in blue.

$E_{DW}^i(\mathbf{k})$ is the energy of the moiré coupled inner wall and $E_{SW}^i(\mathbf{k})$ is the energy of the SWCNT. $|\mathbf{k}_\parallel| = 2\pi/|T|$ is the wave vector along the tube axis, with $T$ being the translational symmetry of the nanotube[12]. The integration is performed for $\mathbf{k} \in (0, \mathbf{k}_\parallel)$ and summation runs over the occupied electronic states. The change in the total energy results in the effective force field $F(R)_i = -\partial \Delta E_i / \partial R$ experienced by the inner tube. We describe the RBM as a harmonic oscillator with frequency $\omega_{SW} = \sqrt{Q_{SW}/M}$ defined by an effective spring constant $Q_{SW}$ and oscillator mass $M$. The force $F(R)_i$ is expanded in a Taylor series over a new equilibrium position $R_0$, $F(R_0)_i + \frac{\partial F(R_0)}{\partial R}(R - R_0) + O(R^2)$ and the equation of motion reads

$$M\ddot{R} = -Q_{SW}(R - R_0) + \frac{\partial F(R_0)}{\partial R}(R - R_0). \qquad (4)$$

Equation 4 implies a new effective spring constant $Q_{DW} = Q_{SW} - \frac{\partial F(R_0)}{\partial R}$ that determines the moiré coupled vibrational frequency $\omega_{DW} = \sqrt{Q_{DW}/M}$, and the approximate shift:

$$\Delta\omega = \omega_{DW} - \omega_{SW} \approx \frac{1}{2M\omega_{SW}} \frac{\partial F(R_0)}{\partial R}. \qquad (5)$$

The force derivative determines the sign of the phonon frequency change. A decrease of the optical transition energies increases the RBM frequencies, whereas an increase corresponds to smaller RBM frequencies. To obtain a relation between the change in $j$-th transition energy $\Delta E_{jj}$ =$(E_{jj}^{DW} - E_{jj}^{SW})$ and the RBM shift $\Delta \omega$ we write the coupling amplitude as $u_0(R) = \gamma_0 e^{R/\lambda} + c$, where $\gamma_0$=8.1 eV, $\lambda$ =0.41 Å, and $c$ =4.2 [7,9]. From Eq. (5) we find

$$\Delta\omega = \frac{2 \sum_m A_m}{m B_j} \frac{\Delta E_{ii}}{M \lambda^2 \omega_{SW}} \approx \alpha_j \Delta E_{jj}. \quad (6)$$

$M$ refers to the mass of two carbon atoms, $A_m$-correspond to the fraction of energy change in the $m$-th electron state, $B_j$ reflects the bands gap variation of the $j$-th optical transition, overall $\alpha_j$ contains inner wall specific parameters, see Supplementary Methods for details. Equation (6) predicts a linear relation between transition energy and frequency shifts in radial phonons like the RBM; we now can verify experimentally this prediction.

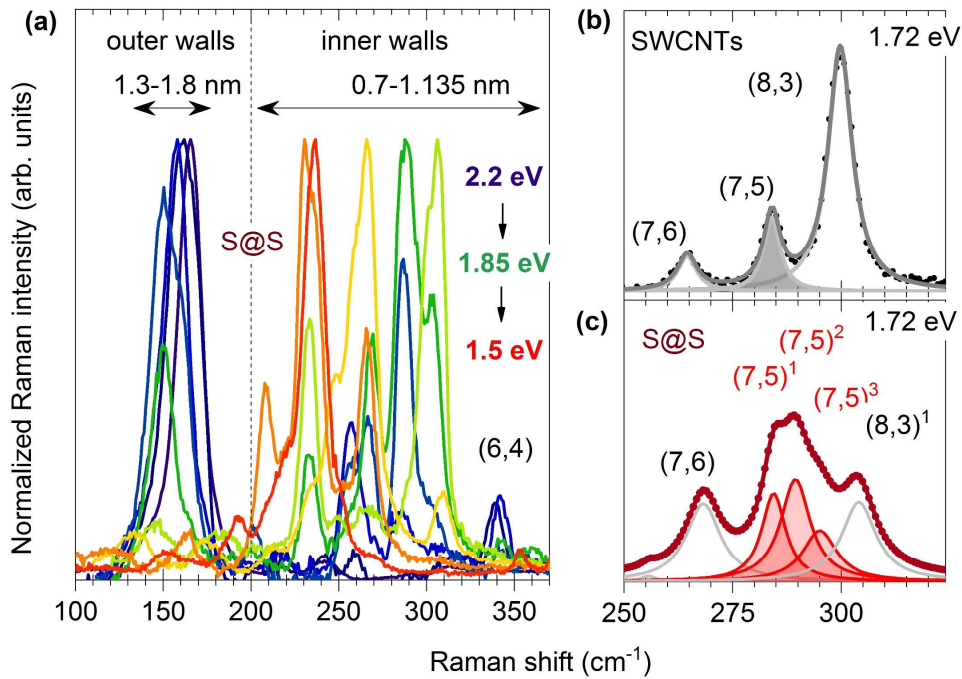

We use resonant Raman spectroscopy to find the RBM frequencies of DWCNTs and their corresponding transition energies $E_{22}$ [12,33,35]. The Raman spectra of a sample of S@S DWCNTs contain RBM peaks of the inner walls for frequencies above 200cm$^{-1}$ and RBMs of the outer walls for smaller frequencies, Fig.2a. Characteristically, the RBM peaks of a nanotube split into several components when the tube is the inner part of a DWCNT as we show for the example of a (7,5) nanotube in Fig.2b and c. By changing the excitation energy, Fig.2a, different tubes get in and out of resonance, which we will use to assign the chiral indices $(n_1^i, n_2^i)$ of inner walls [35,36]. Generally, the RBM peaks from different combinations of inner and outer walls overlap. We, therefore, start our analysis with the (6,4) inner wall, because its RBM is well separated from the other peaks, Fig.2a. We collect the Raman spectra at 15 wavelength across the (6,4) $E_{22}$ resonance, see exemplary spectra in Fig.3a. The peak appears to shift in frequency with laser energy, but this impression is incorrect, instead the RBM region contains at least six different components with constant frequencies between 332-347 cm$^{-1}$. The varying distribution between intensities leads to the impression of an overall frequency shift. Each component corresponds to at least one DWCNT. The frequency of the (6,4) SWCNT (RBM$_{(6,4)}$ = 333 cm$^{-1}$, vertical line) increased by up to 14 cm$^{-1}$ or almost 5% in the DWCNTs.

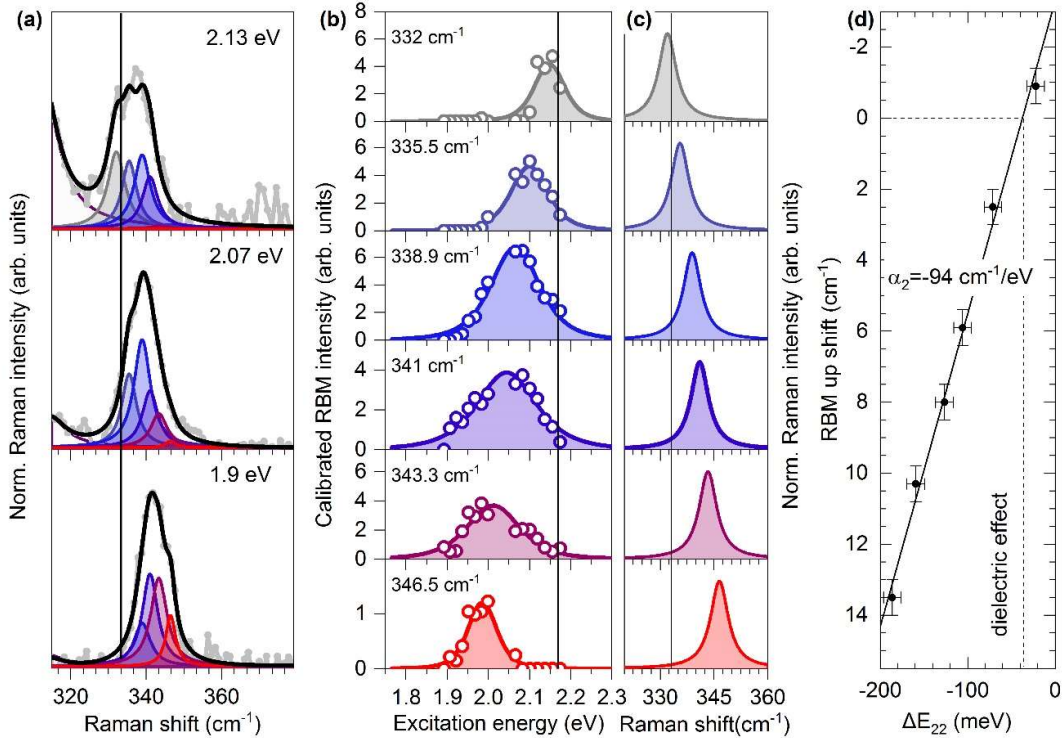

**Figure 3**. Resonant Raman spectroscopy of (6,4)@$(n_1^o, n_2^o)$ DWCNTs, (a) spectral composition of exemplary Raman spectra excited at 2.13, 2.07, and 1.9 eV (from top to bottom). The peak color highlights the different outer tubes. (b) Raman resonance profiles of the RBMs shown in (c), where the integrated peak area is plotted versus excitation energy. The solid vertical line corresponds to the $E_{22}$ transition of the (6,4) SWCNT. Vertical line in (a) and (c) corresponds to the RBM of SWCNT. (d) Difference between the (6,4)@$(n_1^o, n_2^o)$ and (6,4) RBM frequency vs the red-shift of the $E_{22}$ in DWCNTs compared to the (6,4) SWCNT.

The resonance energies of the inner tube $E_{22}$ also differ with varying RBM frequency, Fig. 3a. The intensity distribution between the modes depends on the excitation energy, at 2.1 eV the low frequency modes were resonant, whereas at 1.9 eV the high-frequency modes were the strongest. We quantified the changes in $E_{22}$ by resonant Raman scattering using fully tunable lasers [33,37,38]. Each $(6,4)@(n_1^o, n_2^o)$ component had a different transition energy $E_{22}$, Fig. 3b and 3c. The shift in resonance between the DWCNTs and SWCNTs $\Delta E_{22}$ = 20-200 meV increased linearly with the change in RBM frequency as shown in Fig. 3d, with the slope of the line $\alpha_2$ = -94 cm$^{-1}$/eV. The experimental data thus fully agree with the predictions of Eq. (6). We note that $\Delta E_{22}$ remains finite at the zero RBM shift. This offset is due to the dielectric effect, because the outer wall induces higher screening compared to an uncovered SWCNTs. The dielectric redshift of ~35 meV lies within theoretical predictions[39].

Equation (6) was used to perform a full assignment of the DWCNTs under investigation and to identify the host chiralities for the (6,4) inner-wall. We performed band structure calculations for all possible hosts by Eq. (1), same as we did in Fig 1d, and calculated the theoretical bandgaps $E_{22}^{DW}$. The differences $\Delta E_{22}$ between $E_{22}^{DW} - E_{22}^{SW}$ were plotted as a function of diameter spacing in Fig. 4. This simulation if for DWCNT of identical inner and outer handedness that appear to grow preferentially[40]. In Suppl. Fig. S3 we show the additional shift for opposite handedness. In most cases it is very small, but, e.g., the (6,4)@(8,13) has an additional shift $\Delta E_{22}$~ -20 meV compared to the (6,4)@(13,8). The horizontal lines show the experimental data, where overlap with theoretical predictions identifies the host chirality. This way the RBM at 346.5 cm$^{-1}$ was assigned to the (6,4)@(12,8) tube, 343.3 cm$^{-1}$ to (6,4)@(17,0), 341 cm$^{-1}$ to (6,4)@(16,2), and 338.9 cm$^{-1}$ to (6,4)@(15,4). The RBM peak at 335.5 cm$^{-1}$ originates from several species including (6,4)@(17,1), (6,4)@(16,3), (6,4)@(14,6), (6,4)@(13,8) and (6,4)@(15,5) and the RBM at 332 cm$^{-1}$ can be attributed to the (6,4)@(18,1), (6,4)@(14,7), and (6,4)@(17,3) nanotubes. The right axis represents vibrational moiré shift, obtained using the experimental $\alpha_2$. In such manner one can predict and RBM frequency and transition energy of any possible (6,4)@$(n_1^o, n_2^o)$ chiral combination.

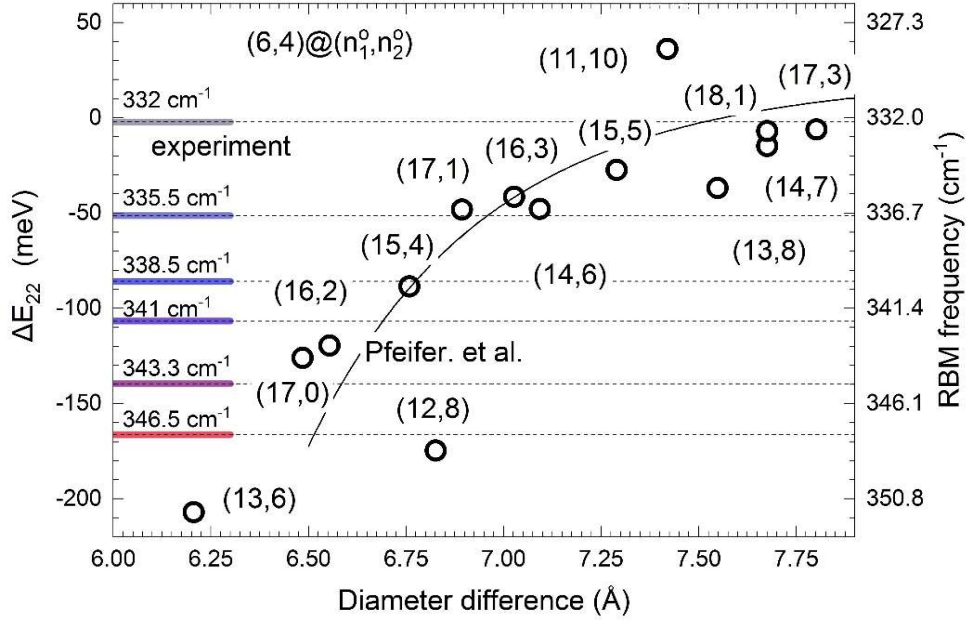

**Figure 4.** Electromechanical coupling in $(6,4)@(n_1^o, n_2^o)$ DWCNTs. The symbols show the calculated with Eq. (1) $E_{22}$ shift compared to the SWCNT, namely $\Delta E_{22} = E_{22}^{DW} - E_{22}^{SW}$ versus diameter difference between the walls. Horizontal lines show experimental data deduced from resonance Raman profiles shifted by dielectric screening (35 meV). The right axis stands for vibrational coupling i.e. RBM shift obtained with Eq. (6) and $\alpha_2$ = -93.8 cm$^{-1}$/eV. The solid line shows the model of Pfeifer et al. [44], see Eq. (S6).

We performed a similar analysis as for the (6,4) inner tube for the eight other inner nanotube species. For these inner walls the RBMs sometimes overlap as shown in Figure 2c. We were able to decompose them experimentally using Eq. (6) under the constrain of using the minimum number of components with constant frequencies. Each fitted peak showed a linear relation between the RBMs and E$_{22}$ shifts, as already observer for the (6,4) species. We show in Figure 5a,b two examples of resonance Raman profiles for (7,5)@$(n_1^o, n_2^o)$ and (9,4)@$(n_1^o, n_2^o)$ nanotubes. We determined the vibrational and transition energies for each sub-peak. Figure 5c shows a Kataura plot of DWCNTs (closed symbols) and SWCNTs (open), i.e., a plot of E$_{22}$ versus $\omega_{RBM}$. For each SWCNT inner wall the corresponding series of DWCNTs forms a line on the Kataura plot. This plot can be used to identify chiral indices and coupling strength via Raman scattering. The $E_{22}$ transition energies of most DWCNTs are smaller than in the corresponding SWCNTs even for little or no shift of the RBMs. As in Figure 3e, we attribute this offset to the dielectric screening of the inner tube by the outer species[41]. The dielectric shift varies between 35 meV in $(6,4)@(n_1^o, n_2^o)$ and 55 meV in $(8,3)@(n_1^o, n_2^o)$ DWCNTs, Fig. 5c. This variation can be explained by a diameter dependence of the self-screening effect[39]. As the tube diameter increases a larger fraction of the exciton wave function extends to the outside resulting in a higher sensitivity to environmental effects [39].

The change in RBM and $E_{22}$ energies are both proportional to the moiré coupling strength. Moiré systems reach strong coupling if the shift of the transition energy $\Delta E_{22}$ is larger than its decay rate $\gamma$. Since the coupling affects both electronic and vibrational states the shift in RBM frequency can be used to identify strong coupling. The typical full width at half maximum (fwhm) at room temperature is determined from the resonant Raman profiles $\gamma=75$ meV. With the experimental $\alpha_2$ we conclude that $\Delta\omega \sim 5$ cm$^{-1}$ signals strong coupling. The red line in Fig. 5d divides strongly and weakly coupled DWCNTs. The majority of the DWCNTs in the S@S sample is in the strong coupling regime.

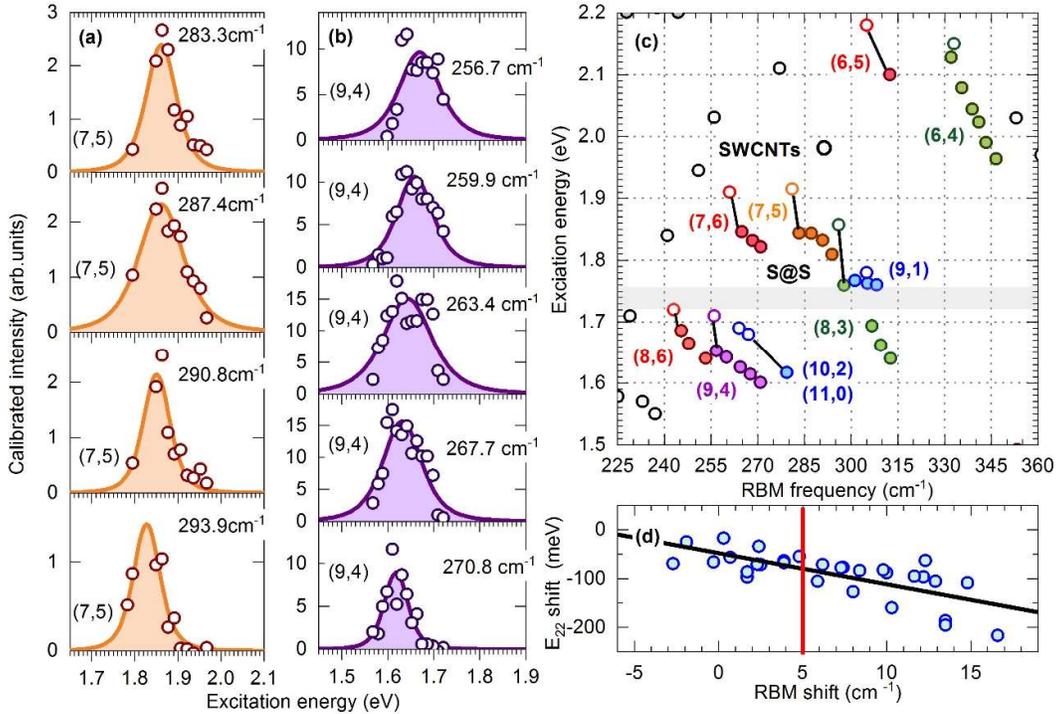

**Figure 5.** Mechanical coupling in small diameter DWCNTs. Resonance Raman profiles of (a) $(7,5)@(n_1^o, n_2^o)$ and (b) for $(9,4)@(n_1^o, n_2^o)$, symbols show the experimental data and lines are the fits. (c) Kataura plot for S@S sample (filled circles) with transition energy $E_{22}$ plotted over RBM frequency, SWCNTs [33] are shown by open circles, different colors were used to separate different inner walls. The values are listed in Suppl. Table S1. The grey line marks the inaccessible in our setup wavelength range. (d) Scattering plot for the $E_{22}$ and RBM shifts, red line divides strongly and weakly coupled DWCNTs.

In the strong coupling regime, the electronic states shift by up to 10% to smaller energies. This regime is particularly interesting for optical studies such as interwall energy transfer, interwall optical excitations, and inelastic scattering. To find such tube system one should aim at DWCNTs with a large shift of the RBM frequency compared to the SWCNT. This will simplify the search of the moiré coupled structures, up to now one required transmission electron microscope to find strongly moiré coupled DWCNTs, like (12,11)@(17,16) found by *Zhao et al.*[9]. The second type of DWCNTs are tubes with flat bands that were predicted for $\varphi \sim 0$ and $\theta \sim \frac{\pi}{6}$ [7]. These tubes are expected to show peculiar transport phenomena such as phase transitions between insulating and superconducting states. Switching between the phases could be exploited for perfect field-effect transistors. Examples of such tubes in the small diameter regime are the $(9,1)@(n_1^o, n_2^o)$ and $(11,0)@(n_1^o, n_2^o)$, see blue symbols in Fig. 5c. Finally, there is the third weakly interacting case with polycrystalline intratube pattern (large $\varphi$ and $\theta$) and large interlayer distance, see Fig 1c. These DWCNTs are suitable for environmental shielding and sheathing in electrical applications [42]. The interaction between the walls is expected to be very weak, which may be the reason for blocked energy transfer between the walls in such DWCNTs [43].

We presented a new model for the RBM shifts in DWCNTs accounting for crystallographic orientations and interlayer distances. The previous model by *Pfeiffer et al.* [21,28] only contained the dependence on the intralayer spacing and is insufficient to reproduce all experimentally observed shifts. We compared our data with the previous model in Figure 4. Stronger deviations are found for DWCNTs with strong coupling ($\varphi$ and $\theta$ are close to zero). A good example is the (6,4)@(12,8) nanotube where we predict RBM frequency 346.5 cm$^{-1}$ as opposed to 339 cm$^{-1}$ from previous works[21,44]. Further examples that we assigned differently are (6,4)@(11,10), (6,4)@(13,8), and (6,4)@(14,8). In weak coupling regime both models seem to deliver similar results, however the underlying model of *Pfeiffer et al.* [21,28] is not compatible with the optical energy shifts. The suggested effective pressure should introduce a strain induced transition energy shift in opposite directions for different $mod(3, n_1^o - n_2^o)$ types of inner walls[45]. In contrast we find the same shift direction $(7,5)@(n_1^o, n_2^o)$ and $(7,6)@(n_1^o, n_2^o)$, see Fig. 5c. Therefore, only the moiré model can reproduce the RBM and transition energies shifts, both in weakly and strongly coupled DWCNTs. The correct RBM shift position is required for identification of the chiral indices, which is critical for understanding the physics effects in DWCNTs and their derivatives.

Moiré coupling also affects vibrational frequencies in more complex one-dimensional heterostructures such as DWCNTs with encapsulated carbon chains [46]. The chirality of the inner nanotube was shown to influence the vibrational frequency of the chain[47]. This statement can be expanded upon using our chiral assignment of DWCNTs, where the RBM at 335 cm$^{-1}$ belongs to a weakly coupled $(6,4)@(n_1^o, n_2^o)$ and not to a $(6,5)@(n_1^o, n_2^o)$ [47]. This DWCNT is in the weak coupling condition and corresponds to 1830.5 cm$^{-1}$ frequency of the encapsulated carbon chain. The same (6,4) inner wall but in the strong coupling regime (RBM = 348 cm$^{-1}$), contains a chain with frequency ~1800 cm$^{-1}$. This large 30 cm$^{-1}$ discrepancy shows that the chirality of the outer wall influences the vibrational frequency of the encapsulated carbon chain. It would be extremely interesting to study the transition energies of the carbon chains and their DWCNTs hosts using multi wavelength resonant Raman scattering.

In conclusion, we showed how moiré coupling affects the vibrations and the electronic transitions of DWCNTs. The transition energy shifts to smaller energy and the RBM frequencies increase for smaller interlayer spacing and aligned chiral vectors, as predicted in two moiré coupled graphene cylinders. The linear correlation between the vibrational and optical shifts is explained by a macroscopic theory, treating the RBMs as two oscillators coupled via the moiré potential. We demonstrated a strong effect of moiré coupling on vibrational states, which will assess electronic coupling in the one-dimensional twisted hetero- and homostructures. Our experiments show that DWCNTs are more than the simple sum of two SWCNTs and need to be treated as novel materials.

**Acknowledgements**
This work was supported by the Focus Area NanoScale of Freie Universitaet Berlin. S. R. acknowledges support by the Deutsche Forschungsgemeinschaft under SPP 2244.

**Associated content.**
**Supporting Information The Supporting Information is available free of charge at** http://pubs.acs.org.

Calculation of the electromechanical coupling; Experimental methods, sample purity analysis; Handedness of DWCNTs, vibrational models comparison

# Supplementary information: Moiré-induced Vibrational Coupling in Double Walled Carbon Nanotubes

Georgy Gordeev [†]*, Sören Wasserroth [†, ‡], Han Li [§], Benjamin Flavel [§], & Stephanie Reich [†]

[†] Department of Physics, Freie Universität Berlin, Arnimallee 14, 14195 Berlin, Germany
[‡] Fritz-Haber-Institut der Max-Planck-Gesellschaft, Faradayweg 4-6, 14195 Berlin
[§] Institute of Nanotechnology, Karlsruhe Institute of Technology, 76021 Karlsruhe, Germany

## Content

1. Calculation of the electromechanical coupling
2. Experimental methods, sample purity
3. Handedness of DWCNTs, vibrational models comparison

### 1. Calculation of the electromechanical coupling

***The band structures and vibrational frequency shifts we derived from the graphene zone folding model.*** The Bravais lattice of graphene is represented by the basis vectors $\boldsymbol{a_1} = a_0(1,0), \boldsymbol{a_2} = a_0(1/2, \sqrt{3}/2)$, with $a_0 = 0.246$ nm that result in reciprocal basis vectors $\boldsymbol{a_g} \cdot \boldsymbol{b_l} = 2\pi \delta_{gl}$. The graphene tight-binding Hamiltonian in the nearest neighbor approximation is given in matrix form by

$$H_i(\boldsymbol{k}) = \begin{pmatrix} 0 & \Delta_k^i \\ \Delta_k^{i\,*} & 0 \end{pmatrix} \qquad (S1),$$

where $\Delta_k = \gamma_c \sum_u e^{i\boldsymbol{k}\cdot\boldsymbol{T}_u^i}$ and $\boldsymbol{T}_u^i$ are the vectors connecting the nearest neighbours in the real space.[1], [2] The hopping energy between graphene lattice sites $\gamma_c$ corresponds to an overlap integral of the electron wavefunctions, we use $\gamma_c = 2.7$ eV. To obtain the Hamiltonian $H_o(\boldsymbol{k})$ of the outer tube in the basis of inner tube, the reciprocal space needs to be rotated by $\varphi \angle (\boldsymbol{C}^i, \boldsymbol{C}^o)$ and expanded by $\frac{|\boldsymbol{C}^i|}{|\boldsymbol{C}^o|}$. For the inner and outer walls of different handedness $\varphi = \angle(\boldsymbol{C}^i, \boldsymbol{C}^o) + 2\theta$, where $\theta$ is chiral angle of the outer tube. These transformations are described by two matrices $P$ and $S^{-1}$ which transform the neighbouring vectors as $T_u^0 = T_u^i S P$ and the Hamiltonian yields

$$H_o(\boldsymbol{k}) = \begin{pmatrix} 0 & \Delta_k^o \\ \Delta_k^{o\,*} & 0 \end{pmatrix}. \qquad (S2)$$

Now the quantized wave vector space around the inner tube circumference correctly describes electronic states of both inner and outer tubes. This condition corresponds to $\boldsymbol{k} \cdot \boldsymbol{C}^i = 2\pi m$ and the allowed wave vectors are typically referred to as cutting lines with $m$ labelling the line number[1], [3].

The tight-binding Hamiltonian of the coupled layers can be written in the form of Eq. (1) with diagonals corresponding to the uncoupled system and off-diagonal U elements describing the intralayer coupling. The off-diagonal elements are given by Eq. (2) and contain a radial component $u_0(R)$ and chiral component depending on the moiré vectors $\boldsymbol{G}_l^M = (\mathbb{1} - S^{-1}P^{-1})\boldsymbol{b}_l$ ($l = 1,2$), where $\mathbb{1}$ is the identity matrix. The Hamiltonian in Eq. (1) is numerically solved for eigenvalues for the allowed wave vectors. These solutions correspond to the moiré coupled electronic bands and are depicted in Figure 2d. Now we can calculate the change in transition energy as function of $u_0(R)$ and we find that is follows quadratic dependence

$$E_{ii} = B_i u_0(R)^2, \qquad (S3)$$

where the $B_i$ is a coefficient for i-th transition. Using the definition in Eq. (4) we can find the change in total energy $\Delta E = \sum \Delta E_m$ expressed by the change of the $m$-th cutting line $\Delta E_m = A_m u_0(R)^2$. We write $u_0(R)$ in form $u_0(R) = \gamma_0 e^{R/\lambda + c}$, where $\gamma_0$=8.1 eV, $\lambda$ =0.41 Å, and $c$ =4.2 [4], [5]. The expression for the second partial derivative of the energy is then

$$\frac{\partial^2 \Delta E(R)}{\partial R^2} = \frac{4 u_0(R)^2}{m \lambda^2} \sum_m A_m. \qquad (S4)$$

Here the energy is normalized by the total number of cutting lines. By substituting Eq. (S2) and Eq. (S1) into Eq. (5) we obtain Eq. (6). Note that $\alpha_i$ is a coefficient for a specific transition. We used the normalized energy in Eq. (S2).

2. **Experimental methods, sample purity**

*Experimental part. Separation.* Separated DWCNTs were used in order to limit the potential combinations of inner and outer wall chiralities. DWCNTs (>60% DWCNT purity, lot no. DW27-096) with an average outer-wall diameter of ~2.0 nm were obtained from NanoIntegris and processed as described previously[6]–[8] to obtained highly enriched S@S DWCNTs. In brief, raw powder (10 mg) was suspended in 20 ml of aqueous 1 wt% SC (Sigma-Aldrich) by tip sonication (Weber Ultrasonics, 35 kHz, 16 W in continuous mode) for 1 h. The resulting dispersion was centrifuged at 45,560g (Beckman Optima L-80 XP, SW 40 Ti rotor) for 1 h and the supernatant collected. This suspension was then processed by gel permeation chromatography using the Sephacryl S-200 gel medium (GE Healthcare) in a glass column (45 cm in length with a 2 cm inner diameter) with 1 wt% SDS solution used as the eluent and a constant flow rate of 2 ml.min$^{-1}$. Fractions (3 ml) were collected from the column and any adsorbed species were finally eluted by addition of 1 wt% SC. The resulted fractions roughly separated into the four types of DWCNTs. The fraction containing predominately S@S DWCNTs was further filtered onto a 0.45 μm Nylon membrane (Phenomenex), washed repeatedly with water and acetone to remove any surfactants and dried overnight in an oven at 120 °C. The dry filter cake was then resuspended by sonication for 1 h with 1 mg ml$^{-1}$ poly[(9,9-dioctylfluorenyl-2,7-diyl)-alt-co-(6,6′-{2,2′-bipyridine})] (PFO–BPy) (American Dye Source) in toluene and centrifuged. The purity of the sample was controlled by the absorption spectroscopy, shown in Figure S1. The absorption of metallic species is strongly reduced. The resulting supernatant was highly enriched in S@S DWCNTs and was used for resonant Raman analysis in this work. Raman from the RBM region provide additional pathway to evaluate the sample purity, see Figure S2. The RBM peaks of metallic species are barely observed.

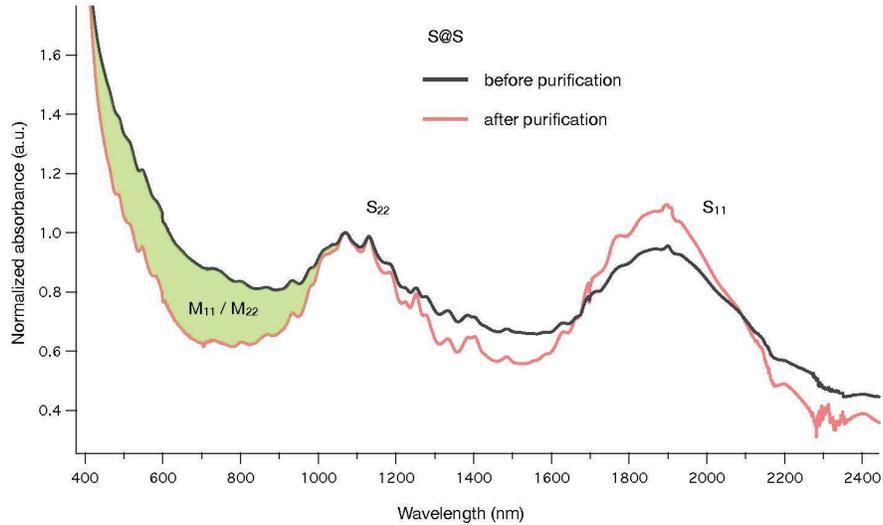

**Figure S1**. Absorption spectroscopy of the S@S sample before (black) and after (red) purification. Green shaded area indicated absorption peaks of metallic species.

***Resonance Raman spectroscopy enables a study of both vibrational states and electronic bandgaps.*** The sorted DWCNT material was pelleted first in order to increase concentration. All samples were centrifuged (Optima Max-XP) at 100,000g for 1 h and dispersed onto a silicon substrate. Raman spectra were measured at the same position on the sample with a tunable laser system. A fluorescent dye laser (Radiant Dyes) with different dyes (R6G: 560–615 nm and DCM: 620–670 nm) and a Coherent Ti:sapphire laser (690–885 nm) were used. The laser power was kept constant at 500 μW to avoid heating the sample. The Raman response of the nanotubes was corrected to a CaF2 reference. The laser light was guided through a ×100 microscope objective and measured with a T64000 Horiba Jobin Yvon spectrometer with a silicon charge-coupled device (CCD) in backscattering configuration. For each *(n,m)* chirality is single walled nanotubes we find the least number Lorentzian peaks with constant positions to fit each single Raman spectra in the full excitation region. The peak area of the RBM was divided by the peak area of the Raman CaF$_2$ reference. This procedure calibrates the intensity by accounting for a wavelength dependent spectrometer sensitivity. The calibrated intensity was then plotted over the excitation energy in Figure 3b and Figure 5a,b. The resulting resonance Raman profiles were fitted with the third-order perturbation theory[9], [10].

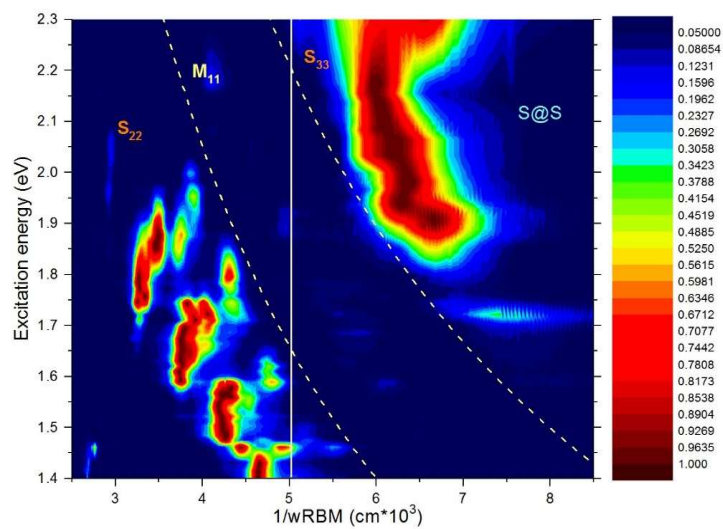

**Figure S2**. Resonant Raman maps of S@S in the region of RBMs excited between 1.5-2 eV. Horizontal axis is inversed RBM, dashed lines divide metallic and semiconducting regions on the Raman map. Vertical line divides inner and outer walls. Each spectrum was normalized to one.

$$I_{RBM}(E_L) \sim \left| \frac{M_{DWCNT}}{\left( E_L - E_{22}^{DW} - \hbar\omega_{Dw} - i\frac{\Gamma}{2} \right)\left( E_L - E_{22}^{DW} - i\frac{\Gamma}{2} \right)} \right|^2, \quad (S5)$$

, where $M_{DWCNT}$ is a complex matrix element containing exciton phonon and exciton-photon interactions, $\Gamma$ is finite lifetime broadening. From the fits, we determine the transition energies and RBM shifts plotted in Figure 3d and Figure 5c. The results of the fit, including optical transition energies are summarized in table S1.

Table S1. RBM fitting parameters, with Eq. (S5) for identified $(n_1^i, n_2^i)$ of the inner wall. The DWCNTs are grouped by laola family index $2n_1^i + n_2^i$. The data for SWCNTs (SW) are taken from Ref. [9], $\hbar\Delta\omega = \hbar\omega_{DW} - \hbar\omega_{SW}$, $\Delta E_{22} = E_{22}^{DW} - E_{22}^{SW}$

| laola family $(n_1^o, n_2^o)$ | $\hbar\omega_{DW}$, cm$^{-1}$ | $E_{22}^{DW}$, eV | $M_{DWCNT}$ | $\frac{\Gamma}{2}$, meV | $E_{22}^{SW}$, eV | $\hbar\omega_{SW}$, cm$^{-1}$ | $\hbar\Delta\omega$, cm$^{-1}$ | $\Delta E_{22}$, meV |
|---|---|---|---|---|---|---|---|---|
| $2n_1^i + n_2^i = 16$ | | | | | | | | |
| (6,4) | 331.1 | 2.12 | 0.008 | 50 | 2.15 | 333 | -1.9 | -25 |
| (6,4) | 335.5 | 2.08 | 0.136 | 77 | 2.15 | 333 | 2.5 | -71 |
| (6,4) | 338.9 | 2.04 | 0.289 | 104 | 2.15 | 333 | 5.9 | -106 |
| (6,4) | 341 | 2.02 | 0.335 | 129 | 2.15 | 333 | 8 | -127 |
| (6,4) | 343.3 | 1.99 | 0.255 | 113 | 2.15 | 333 | 10.3 | -160 |
| (6,4) | 346.5 | 1.96 | 0.037 | 54 | 2.15 | 333 | 13.5 | -186 |
| $2n_1^i + n_2^i = 17$ | | | | | | | | |
| (6,5) | 312.4 | 2.10 | 0.193 | 77 | 2.18 | 305 | 7.4 | -77 |
| $2n_1^i + n_2^i = 20$ | | | | | | | | |
| (7,6) | 264.9 | 1.85 | 0.257 | 85 | 1.91 | 261 | 3.9 | -64 |
| (7,6) | 268.3 | 1.83 | 0.157 | 63 | 1.91 | 261 | 7.3 | -78 |
| (7,6) | 271 | 1.82 | 0.050 | 39 | 1.91 | 261 | 10 | -88 |
| $2n_1^i + n_2^i = 19$ | | | | | | | | |
| (7,5) | 283.3 | 1.84 | 0.182 | 58 | 1.915 | 281 | 2.3 | -71 |
| (7,5) | 287.2 | 1.84 | 0.421 | 92 | 1.915 | 281 | 6.2 | -72 |
| (7,5) | 290.8 | 1.83 | 0.127 | 49 | 1.915 | 281 | 9.8 | -82 |
| (7,5) | 293.9 | 1.81 | 0.103 | 49 | 1.915 | 281 | 12.9 | -106 |
| (8,3) | 297.7 | 1.76 | 0.238 | 46 | 1.857 | 296 | 1.7 | -97 |
| (9,1) | 301.2 | 1.77 | 0.167 | 53 | 1.78 | 305 | -3.8 | -13 |
| (9,1) | 305.3 | 1.76 | 0.246 | 50 | 1.78 | 305 | 0.3 | -17 |
| (9,1) | 306.7 | 1.69 | 0.195 | 66 | 1.78 | 305 | 1.7 | -86 |
| (8,3) | 308.2 | 1.76 | 0.302 | 60 | 1.857 | 296 | 12.2 | -97 |
| (8,3) | 309.5 | 1.66 | 0.048 | 31 | 1.857 | 296 | 13.5 | -195 |
| (8,3) | 312.6 | 1.64 | 0.025 | 24 | 1.857 | 296 | 16.6 | -217 |
| $2n_{1+}^{i} n_2^i 22$ | | | | | | | | |
| (8,6) | 245.4 | 1.69 | 0.164 | 43 | 1.72 | 243 | 2.4 | -34 |
| (8,6) | 247.8 | 1.67 | 0.313 | 59 | 1.72 | 243 | 4.8 | -55 |
| (8,6) | 253.3 | 1.64 | 0.335 | 67 | 1.72 | 243 | 10.3 | -80 |
| (9,4) | 256.7 | 1.65 | 0.619 | 78 | 1.71 | 256 | 0.7 | -57 |
| (9,4) | 259.9 | 1.64 | 0.501 | 68 | 1.71 | 256 | 3.9 | -68 |
| (9,4) | 264.4 | 1.63 | 1.119 | 94 | 1.71 | 256 | 8.4 | -84 |
| (9,4) | 267.65 | 1.61 | 0.701 | 73 | 1.71 | 256 | 11.65 | -95 |
| (9,4) | 270.8 | 1.60 | 0.168 | 39 | 1.71 | 256 | 14.8 | -109 |
| (10,0),(11,0) | 279.3 | 1.62 | 0.132 | 59 | 1.68 | 267 | 12.3 | -63 |

Table S2. Identification of the (6,4) CNT hosts using electromechanical coupling.

| $\omega_{DW}$, cm$^{-1}$ | $E_{ii}^{DW}$, eV | (6,4)@$(n_1^o, n_2^o)$ |
|---|---|---|
| 331.1 | 2.12 | (17,3), (14,7), (18,1), (15,5) |
| 335.5 | 2.08 | (13,8), (14,6), (16,3), (17,1) |
| 338.9 | 2.04 | (15,4) |
| 341 | 2.02 | (16,2) |
| 343.3 | 1.99 | (17,0) |
| 346.5 | 1.96 | (12,8) |

### 3. Handedness of DWCNTs and models comparison

The model of *Koshino et al.* accounts only for the CNTs with the same handedness. An additional rotation by $2\theta$ between the chiral vectors is required in order to account for differently handed inner and outer walls, as described in the first Supplementary section. In order to estimate this effect we have calculated the expected transition energy shifts for (6,4) hosts of different handedness, see Fig. S3. The coupling in the DWCNTs with weak chiral component does not noticeably change, however the coupling in chirally coupled species differ depending on handedness. Despite theoretical calculation, it is still possible that actual tubes grow in the same chiral direction, as shown by only available study [11]. Predicted difference in transition energies and vibrational frequencies provides intriguing possibilities for isomeric studies of DWCNTs.

With Figure S3 directly compares our model with previously used one of *Pfeifer et al.* given by the equation[12]:

$$\hbar\omega_{RBM}(dR) \sim \left(\frac{0.794 \text{ nm}}{2dR}\right)^{14.2} + 330 cm^{-1} \quad (S6)$$

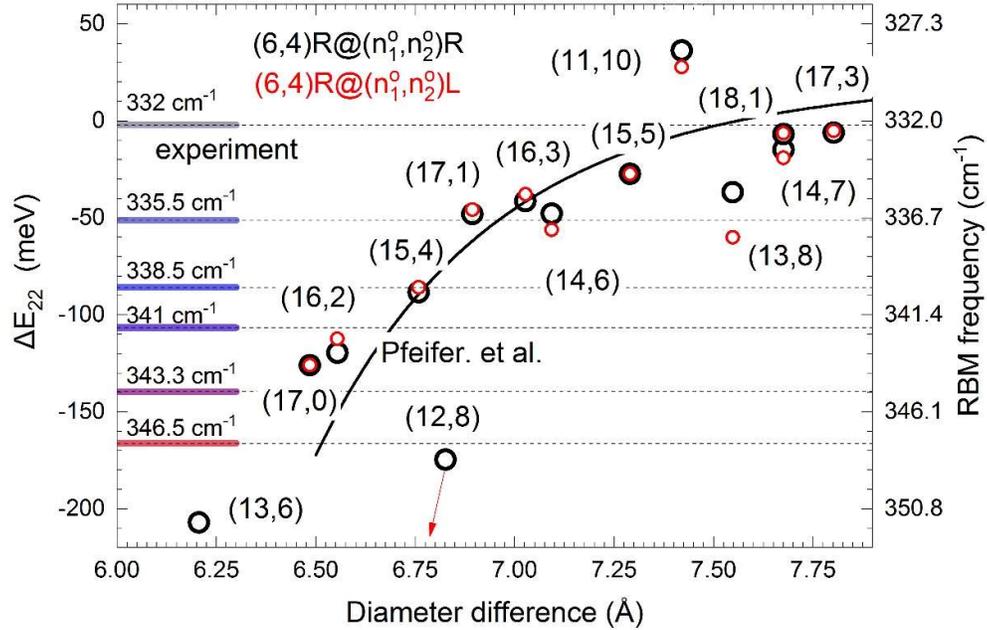

**Figure S3.** Electromechanical coupling in $(6,4)@(n_1^o, n_2^o)$ DWCNTs. The symbols show the calculated with Eq. (1) $E_{22}$ shift compared to the SWCNT, namely $\Delta E_{22} = E_{22}^{DW} - E_{22}^{SW}$ versus diameter difference between the walls. Red (black) symbols represent tubes with the different (same) handedness. Horizontal lines show experimental data deduced from resonance Raman profiles shifted by dielectric screening (35 meV). The right axis stands for vibrational coupling i.e. RBM shift obtained with Eq. (6) and $\alpha_2 = -93.8$ cm$^{-1}$/eV. The black solid line represent the model of Pfeifer et al. [12] given by Eq.(S6).

The base frequency was shifted by 2 cm$^{-1}$ to account for temperature difference. Equation S6 only depends on the diameter differences and chiralities and helicities of the walls are ignored. In contrast, our model gives more elaborated result, the influence of the strong coupling can be seen by the outliers from the line. In strong coupling regime, where the chiral vectors align and the intrawall spacing is small, the moiré based model delivers deviating RBM frequencies. Therefore, now the vibrational frequencies of the following tubes need to be reconsidered (6,4)@(14,7), (6,4)@(13,8), (6,4)@(11,10), (6,4)@(11,9), and (6,4)@(12,8).

**References.**